\documentclass{article}

 \usepackage[preprint]{neurips_2026}

\usepackage[utf8]{inputenc} 
\usepackage[T1]{fontenc}    
\usepackage{hyperref}       
\usepackage{url}            
\usepackage{booktabs}       
\usepackage{amsfonts}       
\usepackage{nicefrac}       
\usepackage{microtype}      
\usepackage{xcolor}         
\usepackage{graphicx}
\usepackage{listings}
\usepackage{multirow}

\title{MermaidSeqBench: An Evaluation Benchmark for NL-to-Mermaid Sequence Diagram Generation}

\author{%
  Basel Shbita \\
  IBM Research \\
  San Jose, CA \\
  \texttt{basel@ibm.com} \\
  \And
  Farhan Ahmed \\
  IBM Research \\
  San Jose, CA \\
  \texttt{farhan.ahmed@ibm.com} \\
  \And
  Chad DeLuca \\
  IBM Research \\
  San Jose, CA \\
  \texttt{delucac@us.ibm.com} \\
}

\begin{document}

\maketitle

\begin{abstract}
Large language models (LLMs) have shown great promise in generating structured diagrams from natural language descriptions, particularly Mermaid sequence diagrams for software engineering. 
However, the lack of existing benchmarks to assess the LLM's correctness on this task hinders rigorous, systematic evaluation and principled comparison of model capabilities on this task. 
To address this shortcoming, we introduce \textbf{MermaidSeqBench}, a human-verified and synthetically extended benchmark for assessing LLM capabilities in generating Mermaid sequence diagrams from natural language prompts. 
The benchmark consists of 132 samples developed via a hybrid methodology of human-verified flows, LLM-based augmentation, and rule-based expansion. 
The evaluation uses an LLM-as-a-judge model to assess generation across various fine-grained metrics such as syntax correctness, activation handling, error handling, and practical usability. 
To demonstrate the effectiveness and flexibility of our benchmark, we perform initial evaluations on numerous state-of-the-art LLMs with multiple LLM judges which reveal significant capability gaps across models and evaluation modes. 
MermaidSeqBench provides a foundation for evaluating structured diagram generation and advances the scientific understanding of LLM capabilities and limitations in structured generation tasks.
\end{abstract}

\section{Introduction}

\begin{figure}[t]
    \centering
    \begin{minipage}{0.48\textwidth}
        \centering
        \includegraphics[width=\linewidth]{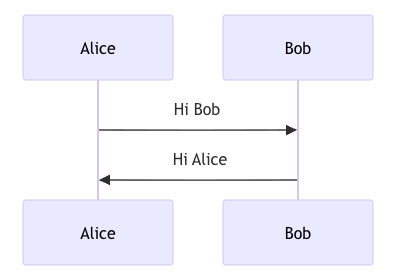}
    \end{minipage}
    \hfill 
    \begin{minipage}{0.48\textwidth}
        \begin{lstlisting}[
            basicstyle=\ttfamily\scriptsize,
            frame=single,
            numbers=none,
            breaklines=true,
            showstringspaces=false,
            xleftmargin=2pt,
            xrightmargin=2pt
        ]
sequenceDiagram
    participant Alice
    participant Bob
    Alice->>Bob: Hi Bob
    Bob->>Alice: Hi Alice
        \end{lstlisting}
    \end{minipage}
    
    \caption{A sample sequence diagram visually represented in UML (a UML sequence diagram) and the corresponding Mermaid code to textually represent it (a Mermaid sequence diagram).}
    \label{fig:mermaid-sample}
\end{figure}

Large language models (LLMs) have demonstrated remarkable capabilities in programming tasks such as code generation \citep{jiang2024surveylargelanguagemodels, huynh2025largelanguagemodelscode}, code documentation \citep{chakrabarty2024freecustomizablecodedocumentation, dvivedi2024comparativeanalysislargelanguage}, and coding assistants \citep{li2025coding}. 
Furthermore, they show great promise in generating structured diagrams from natural language descriptions \citep{guernsey2025harnessing}.
One type of structured diagram is a sequence diagram: a visual model used in software engineering that shows how objects, components, and processes interact with each other over time. 
Sequence diagrams are most commonly visualized in Unified Modeling Language (UML) \citep{uml251}, but can also be represented in a text-based syntax such as PlantUML \citep{plantuml} or Mermaid \citep{mermaid}.
A sample sequence diagram visually represented in UML and textually in Mermaid is shown in Figure~\ref{fig:mermaid-sample}. 

LLMs have proven to be proficient in generating such text-based sequence diagrams \citep{munima2021generatingsequence, wang2024llmsaidumlmodeling}.
However, systematic evaluations for assessing an LLM's correctness in producing sequence diagrams remains largely underdeveloped. 
This has significant implications for software engineering since sequence diagrams are critical for modeling how components interact over time.
Without a systematic benchmark, it is impossible to rigorously characterize an LLM’s strengths and weaknesses on this task, or to make principled comparisons across models and evaluation criteria. 
This has motivated research on developing evaluations for sequence diagram generation such as statistical method~\citep{ferrari2024modelgenerationllmsrequirements} or an LLM-as-a-Judge~\citep{ahmed2025mcetbehavioralmodelcorrectness}.

Mermaid \citep{mermaid}, formally known as MermaidJS, is a diagramming tool that uses a Markdown-inspired syntax and is one such text-based method of representing sequence diagrams. 
Sequence diagrams textually represented in Mermaid are aptly named Mermaid sequence diagrams. 
Prior work has shown that LLMs are capable of effectively producing Mermaid sequence diagrams \citep{saxena2025dynamic}.
However, evaluation in this space is even more underexplored than evaluation for PlantUML and UML sequence diagrams, as there is a severe lack of existing benchmarks or even public datasets to evaluate an LLM's Mermaid sequence diagram generation capabilities. 
Without such assessments, it is impossible to make principled claims about model capability or to identify where models succeed and fail on this task.
This gap motivates our work to create a reproducible, scalable, and fine-grained evaluation framework for Mermaid sequence diagram generation.

In this paper, we introduce \textbf{MermaidSeqBench}, a human-verified and synthetically extended benchmark for assessing an LLM's capabilities in generating Mermaid sequence diagrams from natural language prompts. This benchmark at its core is a dataset that consists of 132 samples that started from a small, manually crafted set and were synthetically expanded via human annotation, in-context LLM prompting, and rule-based variation generation. We also introduce a systematic evaluation method using an LLM-as-a-Judge to assess Mermaid sequence diagram generation across fine-grained metrics such as syntax correctness, activation handling, error handling, and practical usability. This allows us to benchmark and numerically evaluate any LLM on its Mermaid sequence diagram generation capabilities.

To demonstrate the effectiveness of MermaidSeqBench, we perform initial evaluations on three distinct families of models of similar sizes: \textit{Llama-3.1-8B-Instruct} \citep{grattafiori2024llama3herdmodels}, \textit{Qwen-2.5-7B-Instruct} \citep{qwen2025qwen25technicalreport}, and \textit{Granite-3.3-8B-Instruct} \citep{granite2024granite}.
We then extend to models of the same families at smaller sizes to demonstrate the scaling effects and trends across different model sizes.
Lastly, to isolate and study scaling effects, we conduct an ablation over multiple sizes of the \textit{Qwen-2.5} instruction-tuned models which allows us to control architectural differences and focus solely on the model scale.
For all of our evaluations, we use two large LLM judge models: \textit{DeepSeek-V3}~\citep{deepseekai2025deepseekv3technicalreport} at 671B and \textit{GPT-OSS}~\citep{openai2025gptoss120b} at 120B.
This highlights both the effectiveness and flexibility of our proposed benchmark as our results reveal significant capability gaps across models and the various evaluation modes.

Our benchmark aims to provide a foundation for advancing research, not only in Mermaid sequence diagram generation, but also structured diagram generation as a whole. As a result, we open-source MermaidSeqBench\footnote{Code available at: \url{https://github.com/IBM/MermaidSeqBench-Eval}} and make the benchmark dataset publicly accessible\footnote{Dataset available at: \url{https://huggingface.co/datasets/ibm-research/MermaidSeqBench}} in hopes of encouraging further work in developing more rigorous, fine-grained evaluation methodologies in this space. Furthermore, although we focus on NL-to-Mermaid sequence diagram evaluations, our methodology can be applied to other structural diagrams or even other textual representations such as PlantUML.

\section{Related Work}

Existing work on evaluating LLM generation of Mermaid sequence diagrams remains largely underexplored and limited. Expanding to other forms of structural diagrams, \citet{saxena2025dynamic} and \citet{guernsey2025harnessing} both explore methods on using LLMs to generate various structural diagrams in Mermaid syntax such as class, flow, and sequence diagrams. However, evaluations remain relatively simple through simple compliance or visualization checks which are not as robust or scalable.

In the broader space, structural diagrams generated by LLMs are typically represented textually in PlantUML \citep{plantuml}. 
There are several works on LLM generation of PlantUML class, flow, case, and sequence diagrams \citep{guernsey2025harnessing, wang2024llmsaidumlmodeling, rouabhia2025behavioralaugmentationumlclass, Eklund1983182}. Evaluations for these tasks provide more robust approaches. \citet{rouabhia2025behavioralaugmentationumlclass} use statistical methods to validate the syntactic, structural, and behavioral consistency of the LLM generated UML class diagrams in similar way to existing code benchmarks such as HumanEval \citep{chen2021codex} or MBPP \citep{austin2021programsynthesislargelanguage}. \citet{ferrari2024modelgenerationllmsrequirements} utilize similar statistical methods to evaluate LLM generated UML sequence diagrams on categories such as completeness, correctness, and adherence to standards. Furthermore, \citet{ahmed2025mcetbehavioralmodelcorrectness} explore using an LLM-as-a-Judge for evaluating the NL-to-PlantUML sequence diagram task.

The use of LLMs as automated evaluators has been established as a scalable alternative to human annotation for open-ended generation tasks~\citep{zheng2023judgingllm}. Recent work has demonstrated its applicability to code generation and structured output evaluation~\citep{thakur-etal-2025-judging, son2024llmasajudgerewardmodel}, though a recurring challenge is the dependence of absolute scores on the specific judge model used. MermaidSeqBench addresses this directly by adopting a judge-agnostic design, validating its conclusions across two structurally distinct judge models, and recommending that results be interpreted in terms of relative model rankings rather than absolute scores.

To the best of our knowledge, no formal benchmark or even public dataset exists for evaluating an LLM's capabilities on producing Mermaid sequence diagrams. Our work is the first to introduce such a benchmark, public dataset, and systematic evaluation framework for this task in a manner analogous to established code generation benchmarks.

\section{Methodology}

MermaidSeqBench provides the entire evaluation pipeline for benchmarking an LLM's capabilities in generating Mermaid sequence diagrams. This comprises both a dataset for generating outputs and a systematic evaluation method involving an LLM-as-a-Judge for assessing the generations. We describe how we constructed the dataset in Section~\ref{sec:dataset} and the evaluation method in Section~\ref{sec:evaluation}.

\subsection{Dataset Construction}
\label{sec:dataset}

\begin{figure*}[ht]
    \centering
    \includegraphics[width=\textwidth]{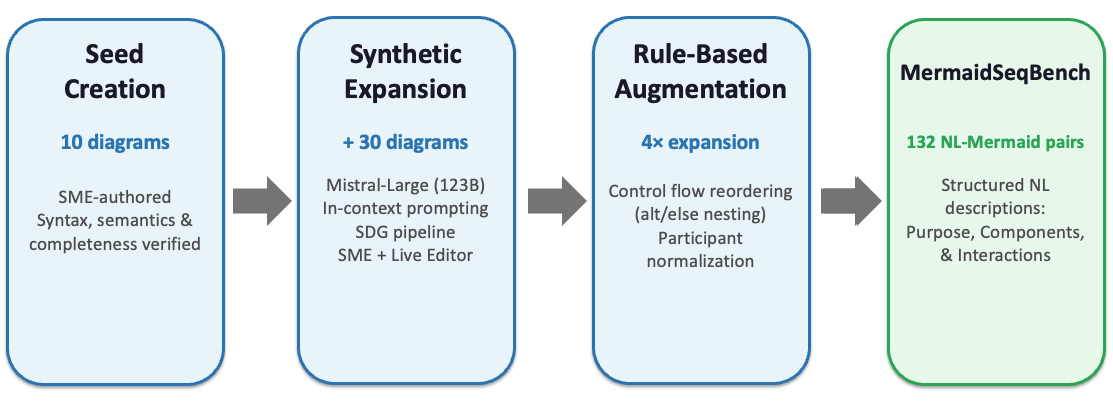}
    \caption{Overview of the MermaidSeqBench dataset construction pipeline. Starting from 10 SME-authored seed diagrams, the benchmark is expanded to 40 verified samples via LLM-based synthetic generation, then scaled to 132 NL-Mermaid pairs through rule-based augmentation. Each diagram is paired with a structured natural language description capturing its purpose, components, and interactions.}
    \label{fig:pipeline}
\end{figure*}

The dataset consists of 132 samples, starting from a small set of manually crafted and verified flows. These were expanded via a hybrid method of combining human annotation, in-context LLM prompting, and rule-based variation generation.
Our methodology for constructing the dataset proceeds in three stages, illustrated in Figure~\ref{fig:pipeline}: (1) initial dataset construction; (2) synthetic expansion using scalable LLM-based generation; and (3) systematic augmentation through rule-based variation.
This pipeline enables us to create a benchmark that is both human-verified at its core and systematically extended for scalability and diversity. See Appendix~\ref{sec:sequence-diagram} for some samples of Mermaid sequence diagrams from the benchmark dataset along with the corresponding UML visualization.

\paragraph{Initial Dataset Seeds}
We begin by constructing a small, high-quality seed set of ten Mermaid sequence diagrams.
These were written manually by a subject matter expert (SME) given a set of natural language descriptions of different sequence flows.
Each flow was manually verified to confirm syntax, semantic plausibility, and completeness.
This curated core serves as the foundation of the benchmark.


\paragraph{Synthetic Expansion}
To scale beyond the core set, we employ \textit{Scalable Synthetic Data Generation (SDG)}~\citep{fms-dgt}, an open-source general-purpose LLM-wrapper framework designed for producing large quantities of synthetic data.
In this stage, we use in-context examples of full flows of valid sequence diagrams in Mermaid syntax to guide the LLM in generating additional diagram flows.
This allows us to systematically extend the benchmark while maintaining fidelity to the Mermaid syntax.
The \textit{SDG} pipeline balances automation with control by leveraging prompts that encourage diverse structural patterns while adhering to syntactic rules. 
We used \textit{Mistral-Large-Instruct} \citep{mistral-large-instruct-2411} (123B) as the primary generator. 
From the generated pool, 30 samples were selected and verified through two complementary mechanisms: (1) manual rendering via the Mermaid Live Editor \citep{mermaid-live-editor} to ensure further syntactic correctness; and (2) manual verification with an SME to confirm completeness and adherence to constraints.


\paragraph{Rule-Based Variation Augmentation}
Finally, we expand the resulting set through deterministic augmentation rules that produce approximately a four-fold increase in coverage.
These rule-based transformations systematically alter both control flow structures and participant naming conventions, ensuring diversity in syntax without altering underlying semantics.
In the first case, conditional constructs defined with \texttt{alt}, \texttt{else}, and \texttt{end} are programmatically detected and reordered, including support for nested alternatives.
This reordering preserves the logical meaning of the branches while producing structurally distinct representations of the same flow.
In the second case, participant identifiers are normalized into a canonical form, with substitutions consistently propagated across both the participant declarations and all subsequent message references.
This guarantees syntactic validity while introducing surface-level variation in the representation of diagram entities.


\paragraph{Natural Language Descriptions}
Each Mermaid diagram in the benchmark is paired with a structured natural language (NL) description capturing its \textit{Purpose}, \textit{Main Components}, and \textit{Interactions}.
This systematic format ensures that every diagram is grounded in a clear, unambiguous textual description, aligning with how interaction flows are typically documented in software design or requirements specifications.
Together, these pairings yield a benchmark of 132 NL-Mermaid pairs in total.
Examples of Mermaid syntax and rendered sequence diagrams can be found in Appendix~\ref{sec:sequence-diagram}, while their corresponding natural language descriptions are detailed in Appendix~\ref{sec:nlp-descriptions}.


\subsection{Dataset Statistics}
\label{sec:dataset-stats}

Table~\ref{tab:dataset_stats} summarizes the syntactic feature distribution across the 132 benchmark samples.
Conditional branching (\texttt{alt}/\texttt{else}) is present in 89.4\% of samples, ensuring robust coverage of multi-branch logic.
Nested control flow (where one \texttt{alt} block is nested within another) appears in 33.3\% of samples, probing more complex interaction patterns.
Activation and deactivation blocks are present in all samples (100\%), which directly motivates the \textit{Activation Handling} evaluation criterion.
Both solid (\texttt{->{}>}) and response (\texttt{-{}->{}>}) message types are well represented.
Participant counts range from 4 to 7 per diagram with an average of 5.0, reflecting realistic multi-component interaction scenarios.
Certain Mermaid constructs beyond conditional branching are not currently represented; this is an acknowledged scope limitation discussed in Section~\ref{sec:limitations}.

\begin{table}[t]
    \centering
    \caption{Distribution of Mermaid syntactic features across the MermaidSeqBench benchmark samples, organized by feature category. Count indicates the number of samples containing each feature, and coverage reflects the corresponding percentage.}
    \label{tab:dataset_stats}
    \resizebox{.75\linewidth}{!}{%
    \begin{tabular}{llrr}
    \toprule
    \textbf{Category} & \textbf{Feature} & \textbf{Count} & \textbf{Coverage (\%)} \\
    \midrule
    \multirow{3}{*}{Control Flow}
        & Conditional branching (\texttt{alt}) & 118 & 89.4 \\
        & Multi-branch (\texttt{else})         & 116 & 87.9 \\
        & Nested control flow                  &  44 & 33.3 \\
    \midrule
    Diagram Elements
        & Activation/deactivation              & 132 & 100.0 \\
    \midrule
    \multirow{2}{*}{Message Types}
        & Solid arrows (\texttt{->{}>})        & 132 & 100.0 \\
        & Response arrows (\texttt{-{}->{}>})  & 126 &  95.5 \\
    \bottomrule
    \end{tabular}
    }
\end{table}


By combining human-verified samples, synthetically expanded instances, and rule-based variations, each paired with a structured natural language description, MermaidSeqBench provides systematic coverage of core Mermaid syntactic features (as quantified in Table~\ref{tab:dataset_stats}) and exposes LLM limitations across a range of control flow structures and interaction patterns.


\subsection{Evaluation Method}
\label{sec:evaluation}

The evaluation pipeline is a two step procedure: (1) the LLM is given the input prompt to generate Mermaid code; and (2) a secondary judge LLM is used to assess the generated Mermaid code based on the task, description, prompt, and expected ``ground truth'' answer.

\paragraph{Inference on Input Prompts}
The evaluation begins by performing model inference using the input prompts from the benchmark dataset. Each sample is formatted in a fixed schema composed of a task directive, general guidelines, Mermaid syntax rules, and example input, an example output, purpose, main components, and interactions. The LLM is expected to produce a valid Mermaid sequence diagram.
To ensure reproducibility, all outputs are generated using greedy decoding (temperature $= 0$) with a maximum budget of 1,024 new tokens per prompt.

\paragraph{LLM-as-a-Judge Assessments}
MermaidSeqBench uses an LLM-as-a-Judge to assess generated Mermaid outputs. Any judge LLM can be used, but larger models are recommended as they tend to be more capable \citep{zheng2023judgingllm, thakur-etal-2025-judging, son2024llmasajudgerewardmodel}. Since only the outputs of the judge model are needed, black-box API models may also be used. Given an input description (in natural language) and a corresponding reference diagram syntax, the judges rate candidate diagrams across six dimensions: \textit{Syntax}, \textit{Mermaid Only}, \textit{Logic}, \textit{Completeness}, \textit{Activation Handling}, and \textit{Error \& Status Tracking}. Each dimension is scored on a scale from 0 to 1, allowing both fine-grained and aggregate comparisons of model performance.

All judge prompts follow a common structure: they present the generated Mermaid diagram (\texttt{<AGENT\_RESPONSE>}) and reference one accepted solution (\texttt{<EXPECTED\_AGENT\_RESPONSE>}) without requiring a verbatim match.
The judge prompt can optionally include the original agent prompt (\texttt{<AGENT\_PROMPT>}).
It also poses targeted questions specific to the evaluation criterion.
Judges are instructed to output both a numerical score in the range $[0.0, 1.0]$ and a brief explanation, using a standardized scoring rubric to promote consistency across criteria and models.
The exact prompts used by the judge models for each of the six criteria are shown in Appendix~\ref{sec:judge-prompts}.

\section{Experimental Setup}
\label{sec:exp_setup}

Our experiments fall into two categories: (1) cross-family and cross-scale comparisons where we evaluate on three different families of models at two similar scales each, and (2) intra-family scaling ablation where we evaluate the same model family across multiple sizes.
For both scenarios, we utilize two large models as judges: \textit{DeepSeek-V3} \citep{deepseekai2025deepseekv3technicalreport} (671B) and \textit{GPT-OSS} \citep{openai2025gptoss120b} (120B).
This allows us to mitigate individual model bias and ensure a more robust evaluation by cross-referencing the scoring of two distinct architectural implementations.


\paragraph{Cross-Family and Cross-Scale}
This experiment probes whether MermaidSeqBench can expose capability differences attributable to model architecture and training rather than scale, by evaluating families at approximately matched parameter counts.
Our initial evaluations are on three distinct families of models of similar sizes: \textit{Llama-3.1-8B-Instruct} \citep{grattafiori2024llama3herdmodels}, \textit{Qwen-2.5-7B-Instruct} \citep{qwen2025qwen25technicalreport}, and \textit{Granite-3.3-8B-Instruct} \citep{granite2024granite}.
We then extend to models within the same families, but of smaller sizes: \textit{Llama-3.2-1B-Instruct}, \textit{Qwen-2.5-1.5B-Instruct}, and \textit{Granite-3.3-2B-Instruct}.
This setup provides a balanced perspective on both lightweight and larger-scale variants across LLM families, enabling comparison of architectural differences as well as scaling effects.


\paragraph{Intra-Family Scaling Ablation}
While the cross-family experiment reveals differences across architectures, it conflates architectural and scaling effects, making it difficult to determine whether observed capability gaps stem from model design or simply model size.
This ablation addresses that question directly by holding architecture constant and varying only scale, providing a controlled view of how Mermaid sequence diagram generation capabilities develop with model capacity.
This is particularly relevant for structured generation tasks, where syntactic precision and multi-step control flow may require a minimum capability threshold that smaller models cannot meet.
We select the \textit{Qwen-2.5} family of models \citep{qwen2025qwen25technicalreport} as it offers the broadest range of instruction-tuned checkpoints among open models, spanning 0.5B to 72B parameters.
Specifically, we evaluate \textit{Qwen-2.5} models at 0.5B, 1.5B, 3B, 7B, 14B, 32B, and 72B using the same benchmark, prompts, and LLM-as-a-Judge setup.

\section{Results and Discussion}
\label{sec:results}

We now present our evaluations on the two categories of experiments and demonstrate how MermaidSeqBench has both the effectiveness and flexibility to reveal significant capability gaps across models and evaluation modes.


\subsection{Cross-Family and Cross-Scale Comparisons}
\label{sec:results_cross_family}

Table~\ref{tab:eval_results} shows the benchmark scores of the \textit{Llama}, \textit{Qwen}, and \textit{Granite} families of models for the two variants of sizes.
Overall, the results reveal clear scaling effects across families: larger models within each family consistently outperform their smaller counterparts, an expected result as larger models are typically more performant.
Among the larger models, \textit{Llama} and \textit{Qwen} achieve the highest scores across most criteria, particularly on syntax, logical flow, and activation usage.
\textit{Granite} follows closely but lags slightly on error handling and completeness.

Beyond overall scaling trends, the results also highlight differences in how model families handle specific aspects of diagram generation.
For example, the \textit{Granite} models perform strongly on strict syntactic and \textit{Mermaid Only} criteria, even at the 2B scale, but lag on dimensions such as activation handling.
The \textit{Qwen} models excel at syntax, while \textit{Llama} exhibits a more balanced profile across metrics.
These divergences point to complementary strengths across families and indicate opportunities for hybrid approaches to structured diagram generation.

Overall, these findings confirm that MermaidSeqBench exposes meaningful differences across model families and sizes, and emphasize the need for a systematic, multi-faceted evaluation of LLMs on diagram generation tasks.


\begin{table}[t]
    \centering
    \caption{Model performance on MermaidSeqBench across six criteria. Each row indicates the candidate generation model; column groups indicate the LLM-as-a-Judge model used for evaluation. Scores are reported as percentages (originally in $[0.0, 1.0]$). For all criteria, higher is better.}
    \label{tab:eval_results}
    \resizebox{\linewidth}{!}{%
    \begin{tabular}{l|rrrrrr|rrrrrr}
    \toprule
     & \multicolumn{6}{c}{DeepSeek-V3 (671B)} & \multicolumn{6}{c}{GPT-OSS (120B)} \\
     \cmidrule(lr){2-7} \cmidrule(lr){8-13}
     & \rotatebox{90}{Syntax} & \rotatebox{90}{Mermaid Only} & \rotatebox{90}{Logic} & \rotatebox{90}{Completeness} & \rotatebox{90}{Activation Handling} & \rotatebox{90}{Error \& Status Tracking} &
     \rotatebox{90}{Syntax} & \rotatebox{90}{Mermaid Only} & \rotatebox{90}{Logic} & \rotatebox{90}{Completeness} & \rotatebox{90}{Activation Handling} & \rotatebox{90}{Error \& Status Tracking} \\
    \midrule
    Llama-3.1-8B-Instruct    & 92.01 & 96.36 & 87.35 & 89.43 & 79.17 & 81.82 & 68.89 & 93.85 & 67.71 & 77.50 & 57.99 & 74.90 \\
    Qwen-2.5-7B-Instruct     & 91.29 & 95.98 & 87.23 & 88.90 & 79.55 & 81.70 & 85.97 & 97.05 & 70.56 & 77.37 & 64.04 & 69.81 \\
    Granite-3.3-8B-Instruct  & 86.97 & 94.13 & 83.03 & 83.75 & 74.13 & 76.97 & 65.15 & 88.35 & 58.90 & 65.08 & 47.08 & 63.24 \\
    \midrule
    Llama-3.2-1B-Instruct    & 68.98 & 60.68 & 52.27 & 60.57 & 39.85 & 52.35 & 46.15 & 46.23 & 18.86 & 25.92 & 17.29 & 24.14 \\
    Qwen-2.5-1.5B-Instruct & 61.59 & 64.55 & 47.31 & 51.61 & 46.06 & 47.77 & 56.07 & 85.22 & 30.49 & 40.26 & 33.99 & 41.53 \\
    Granite-3.3-2B-Instruct  & 75.27 & 90.98 & 70.23 & 74.55 & 63.71 & 71.86 & 39.60 & 78.59 & 34.11 & 46.50 & 29.01 & 53.97 \\
    \bottomrule
    \end{tabular}
    }
\end{table}


\subsection{Intra-Family Scaling Ablation Analysis}
\label{sec:results_ablation}

Table~\ref{tab:eval_results_ablation} reports results for the \textit{Qwen-2.5} family across multiple model sizes, revealing a broadly positive scaling trend across all six evaluation criteria.
The most pronounced gains occur between the 3B and 7B variants, particularly for syntax, logic, and completeness, reflecting a significant capability threshold at that scale.
Under \textit{DeepSeek-V3}, the trend is not strictly monotonic: the 14B variant scores slightly below the 7B on several criteria before recovering at 32B and 72B, suggesting that judge calibration interacts with model scale in non-trivial ways.
Under \textit{GPT-OSS}, improvement is more consistently monotonic across sizes.
Larger variants (32B and 72B) approach saturation on syntax and \textit{Mermaid Only} criteria under both judges, while continuing to yield incremental improvements on more semantically demanding dimensions such as error and status tracking.
Overall, these results indicate that scaling within a single architecture substantially reduces capability gaps, and demonstrate that MermaidSeqBench is sensitive to fine-grained performance differences induced by model size.


\begin{table}[t]
    \centering
    \caption{\textit{Qwen-2.5} scaling results on MermaidSeqBench across six criteria. Each row indicates the candidate generation model; column groups indicate the LLM-as-a-Judge model used for evaluation. Scores are reported as percentages (originally in $[0.0, 1.0]$). For all criteria, higher is better.}
    \label{tab:eval_results_ablation}
    \resizebox{\linewidth}{!}{%
    \begin{tabular}{l|rrrrrr|rrrrrr}
    \toprule
     & \multicolumn{6}{c}{DeepSeek-V3 (671B)} & \multicolumn{6}{c}{GPT-OSS (120B)} \\
     \cmidrule(lr){2-7} \cmidrule(lr){8-13}
     & \rotatebox{90}{Syntax} & \rotatebox{90}{Mermaid Only} & \rotatebox{90}{Logic} & \rotatebox{90}{Completeness} & \rotatebox{90}{Activation Handling} & \rotatebox{90}{Error \& Status Tracking} &
     \rotatebox{90}{Syntax} & \rotatebox{90}{Mermaid Only} & \rotatebox{90}{Logic} & \rotatebox{90}{Completeness} & \rotatebox{90}{Activation Handling} & \rotatebox{90}{Error \& Status Tracking} \\
    \midrule
    Qwen-2.5-0.5B-Instruct   & 58.90 & 77.12 & 36.93 & 44.39 & 26.52 & 38.07 & 48.95 & 65.45 & 13.91 & 18.41 & 13.85 & 15.90 \\
    Qwen-2.5-1.5B-Instruct & 61.59 & 64.55 & 47.31 & 51.61 & 46.06 & 47.77 & 56.07 & 85.22 & 30.49 & 40.26 & 33.99 & 41.53 \\
    Qwen-2.5-3B-Instruct   & 62.20 & 69.39 & 60.57 & 63.64 & 55.91 & 58.94 & 66.80 & 92.67 & 46.86 & 55.35 & 49.92 & 58.87 \\
    Qwen-2.5-7B-Instruct     & 91.29 & 95.98 & 87.23 & 88.90 & 79.55 & 81.70 & 85.97 & 97.05 & 70.56 & 77.37 & 64.04 & 69.81 \\
    Qwen-2.5-14B-Instruct     & 83.79 & 88.94 & 84.77 & 85.19 & 72.31 & 70.55 & 80.87 & 97.50 & 76.65 & 80.06 & 63.06 & 70.97 \\
    Qwen-2.5-32B-Instruct  & 87.92 & 89.39 & 86.29 & 86.82 & 75.21 & 72.16 & 90.32 & 97.73 & 78.72 & 83.23 & 80.20 & 70.55 \\
    Qwen-2.5-72B-Instruct     & 88.56 & 90.00 & 86.33 & 86.36 & 75.91 & 72.46 & 87.14 & 97.95 & 83.38 & 87.69 & 80.85 & 70.66 \\
    \bottomrule
    \end{tabular}
    }
\end{table}


\subsection{Cross-Judge Analysis}
\label{sec:cross-judge}

A notable feature of the results is the systematic difference in absolute scores between the two judge models: \textit{DeepSeek-V3} assigns consistently higher scores across all criteria and model families, while \textit{GPT-OSS} scores more strictly, with gaps of up to 24 percentage points on individual criteria.
This magnitude of inter-judge variance is consistent with documented behavior of LLM-as-a-Judge across evaluation settings~\citep{zheng2023judgingllm, thakur-etal-2025-judging}, where absolute score calibration varies substantially across judge models even when relative assessments remain directionally stable.

Despite this absolute divergence, both judges support the same high-level conclusions.
Larger models substantially outperform their smaller counterparts across all three model families regardless of which judge is used.
The intra-family scaling ablation reveals a broadly positive improvement with model size under both judges, with the trend more consistently monotonic under \textit{GPT-OSS} than \textit{DeepSeek-V3}.
Both judges also agree on the relative difficulty of evaluation criteria: \textit{Mermaid Only} and \textit{Syntax} are consistently the strongest dimensions across models, while \textit{Activation Handling} and \textit{Error \& Status Tracking} represent the most challenging.
Fine-grained rankings within groups of similarly-sized models can differ between judges, which further underscores the value of multi-judge evaluation.

These observations motivate a key interpretive guideline for MermaidSeqBench: scores should be used to compare models relative to one another and to identify capability trends, rather than as absolute measures of diagram generation quality.
Because LLM-as-a-Judge calibration is inherently judge-dependent, MermaidSeqBench is designed to be judge-agnostic: any sufficiently capable judge model may be used, and conclusions should be drawn from patterns that are consistent across judges rather than from the magnitude of any individual score.


\subsection{Benchmark Scope and Evaluative Claims}
\label{sec:scope}

MermaidSeqBench is designed to support \textit{comparative} evaluation: its primary claim is that LLMs can be systematically ranked and differentiated on Mermaid sequence diagram generation across well-defined syntactic and semantic dimensions.
The results in Sections~\ref{sec:results_cross_family} and \ref{sec:results_ablation} demonstrate that the benchmark is sensitive to meaningful differences across model families and scales, and that these differences are robust to the choice of judge model at the level of high-level trends (Section~\ref{sec:cross-judge}).
These properties make MermaidSeqBench suitable for answering questions of the form: \textit{which models are more capable at this task, and along which dimensions do they differ?}

These claims rest on three core assumptions.
First, that LLM-as-a-Judge with a sufficiently capable model provides a valid proxy for assessing structured generation quality, an assumption supported by the directional consistency observed across two structurally distinct judge models.
Second, that the six evaluation criteria collectively capture the dimensions of correctness that matter most for this task.
Third, that the dataset's systematic coverage of core Mermaid syntactic features (Table~\ref{tab:dataset_stats}) is sufficient to surface comparative differences between models, even if it does not exhaust the full space of possible diagrams.

The benchmark does not support claims about absolute diagram quality: as shown in Section~\ref{sec:cross-judge}, absolute scores are calibration-dependent and should not be interpreted as ground-truth measures.
Nor does it claim out-of-the-box generalization to other diagram formats or syntaxes.
Beyond these boundaries, the methodology itself (hybrid dataset construction via human seeds, LLM-based synthetic expansion, and rule-based augmentation, paired with a judge-agnostic evaluation pipeline) is transferable to other structured diagram formats and textual representation languages, offering a reusable template for evaluation in this broader space.


\section{Future Work}

Several directions emerge naturally from this work.
First, the dataset currently omits parallel flows, loop structures, and optional blocks (Section~\ref{sec:dataset-stats}); expanding the seed set to cover these constructs would broaden coverage to more complex interaction patterns.
Second, the hybrid construction methodology is directly applicable to other Mermaid diagram types such as flowcharts, class diagrams, and Gantt charts, as well as to other textual representation languages such as PlantUML, enabling systematic cross-format comparison of LLM structured generation capabilities.
Third, establishing a human-annotated gold standard for a subset of the benchmark would enable formal validation of LLM-as-a-Judge quality and provide calibration anchors for comparing judge models, directly addressing the absolute score interpretation challenge identified in Section~\ref{sec:cross-judge}.
Finally, MermaidSeqBench provides a natural testbed for studying the effect of targeted fine-tuning on structured diagram generation, since the fine-grained criteria allow precise diagnosis of which capabilities improve with training and which gaps persist.

\section{Limitations and Broader Impacts}
\label{sec:limitations}

\paragraph{Limitations}
Our proposed benchmark has several limitations and ethical considerations.
First, while the benchmark comprises 132 samples, this scale is comparable to well-established coding benchmarks such as HumanEval \citep{chen2021codex} (164 samples) and LiveBench-Coding \citep{livebench} (128 samples); the dataset is specifically structured to cover the core syntactic features of Mermaid sequence diagrams rather than maximize sample count. That said, the seed test cases may be influenced by inductive bias from subject matter experts, whose preferences may implicitly shape the flows selected and verified, and the benchmark can be readily extended using the methodology described in Section~\ref{sec:dataset}.
Second, the benchmark focuses exclusively on Mermaid sequence diagrams, leaving out other diagram types and alternative syntaxes (e.g., PlantUML), which may limit generalizability. Within Mermaid, certain constructs beyond conditional branching (e.g., parallel flows and loop structures) are not currently represented in the dataset, leaving these interaction patterns as directions for future expansion.
Third, our findings concentrate on our selection of evaluation criteria; although these criteria are grounded and enable us to explore capability gaps across scales, other categories remain unexplored and could yield greater insights.
Finally, the reliance on LLM-as-a-judge and heuristic-based scorers introduces potential biases and inconsistencies, especially in ill-defined evaluation settings where multiple valid solutions may exist.
These factors highlight directions for future work in broadening coverage, scaling to larger models, and refining judge methodologies.

\paragraph{Broader Impacts}
MermaidSeqBench is released as an open benchmark and evaluation framework, contributing a reusable resource for the research community to systematically assess LLM capabilities in structured generation tasks.
By establishing fine-grained evaluation criteria and a judge-agnostic methodology, the work supports more rigorous and reproducible evaluation practices in the broader NLP and AI community.
We foresee no meaningful negative societal impacts from this work: the benchmark evaluates diagram generation in a controlled research setting and presents no meaningful dual-use or misuse potential.

\section{Conclusion}

We introduced MermaidSeqBench, a benchmark for evaluating LLM capabilities in generating precise, structured, and logically consistent Mermaid sequence diagrams.
By combining human-verified flows with rule-based and LLM-driven expansion, we enable a systematic assessment of fine-grained criteria including syntax, logic, completeness, activation handling, and error/status tracking.
Our evaluations on several state-of-the-art LLMs reveal significant performance gaps across model families, scales, and evaluation criteria.
Notably, \textit{Activation Handling} and \textit{Error \& Status Tracking} emerge as consistently the most challenging dimensions across all models and scales, pointing to structured control flow and error representation as key capability gaps in current LLMs.
Our cross-judge analysis further reveals that while absolute scores are judge-dependent, high-level capability trends remain stable across structurally distinct judge models, a finding with broader implications for how LLM-as-a-Judge benchmarks should be designed and interpreted.
These findings underscore the need for specialized benchmarks like MermaidSeqBench to enable systematic, fine-grained evaluation of LLM capabilities in structured generation tasks.
By open-sourcing both the dataset and evaluation code, and providing a judge-agnostic evaluation pipeline that extends naturally to other structured diagram formats, MermaidSeqBench lays a foundation for more principled and reproducible evaluation methodology in this space.

\bibliography{references}
\bibliographystyle{unsrtnat}

\appendix
\section{Sequence Diagrams}
\label{sec:sequence-diagram}

The UML sequence diagram in Figure~\ref{fig:uml-sequence-diagram} and the accompanying Listing~\ref{lst:uml-sequence-mermaid} illustrate one of the test cases included in our benchmark dataset.
The figure presents the rendered Mermaid sequence diagram describing the flow in uploading documents with secure storage, while the listing provides the exact Mermaid syntax used to generate it.

Our benchmark also includes more complex flows with multiple active actors.
Figure~\ref{fig:uml-sequence-diagram-2} presents a sequence diagram demonstrating a user interaction with a Chatbot actor that may or may not invoke an additional actor (i.e., a Customer Support Agent).
Listing~\ref{lst:uml-sequence-mermaid-2} shows the Mermaid syntax corresponding to this diagram, which can be directly rendered to produce the figure. 


\section{Natural Language Descriptions of Sequence Diagrams}
\label{sec:nlp-descriptions}

For each Mermaid diagram provided in Appendix~\ref{sec:sequence-diagram}, we include the corresponding natural language description that served as the input prompt for generation and evaluation.  
Each description systematically covers: (1) \textbf{Purpose:} the overall intent of the sequence diagram; (2) \textbf{Main Components:} the participants involved and their roles; and (3) \textbf{Interactions:} the ordered set of messages and control-flow constructs.

\begin{figure*}[htbp]
    \centering
    \includegraphics[width=\linewidth]{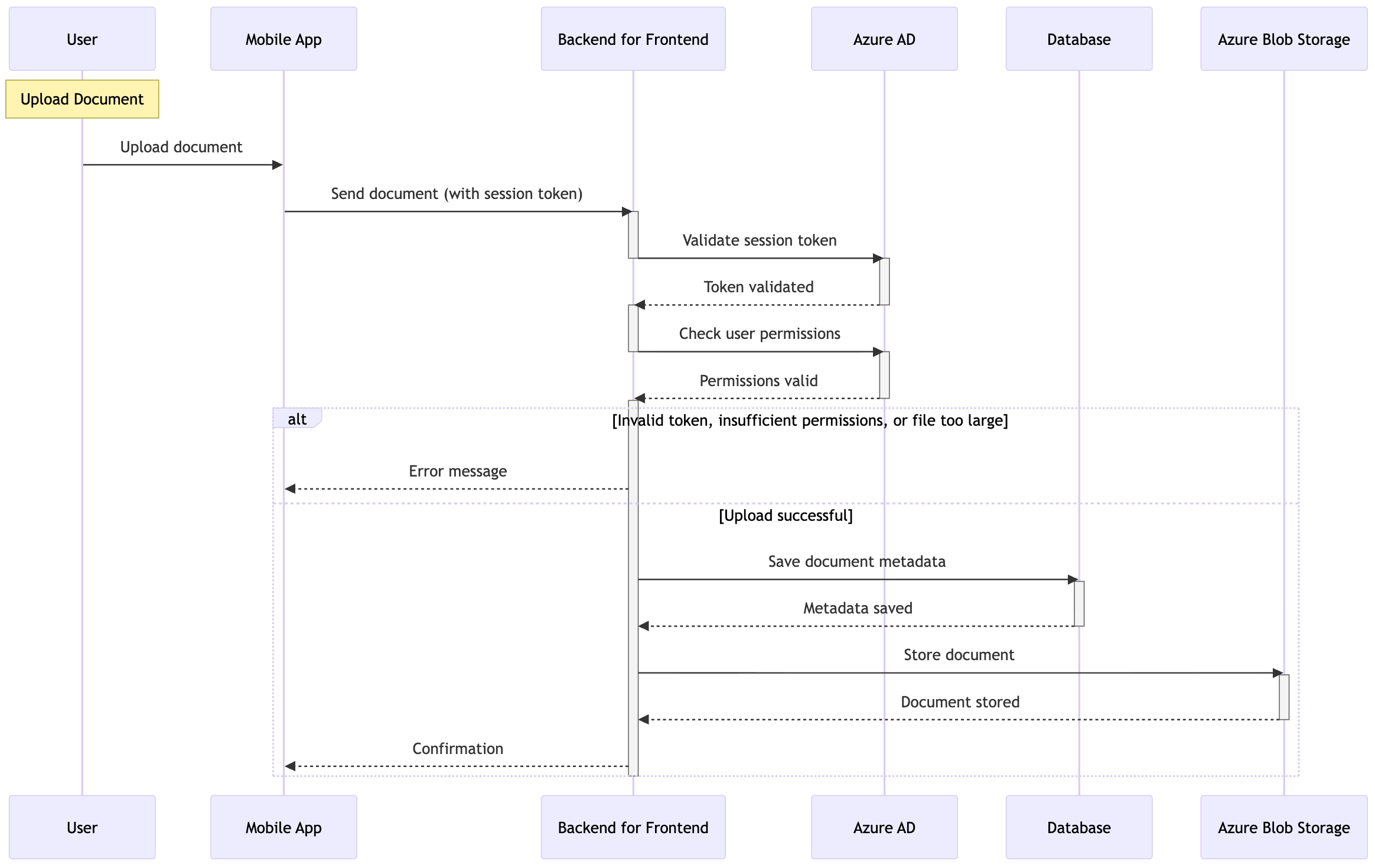}
    \caption{A UML sequence diagram from our benchmark, illustrating the ``Uploading Documents with Secure Storage'' flow. Participants include the User, Mobile App, Backend For Frontend (BFF), Azure AD, Database, and Azure Blob Storage. In this scenario, the User uploads a document through the Mobile App, which forwards the file and session token to the BFF. The BFF validates the token with Azure AD, checks the user's permissions, and, if authorized, records document metadata in the Database and securely stores the file in cloud storage (Azure Blob Storage). A confirmation is then returned to the app, while alternate paths handle errors for unauthorized access or oversized files.}
    \label{fig:uml-sequence-diagram}
\end{figure*}

\begin{figure*}[htbp]
    \centering
    \includegraphics[width=\linewidth]{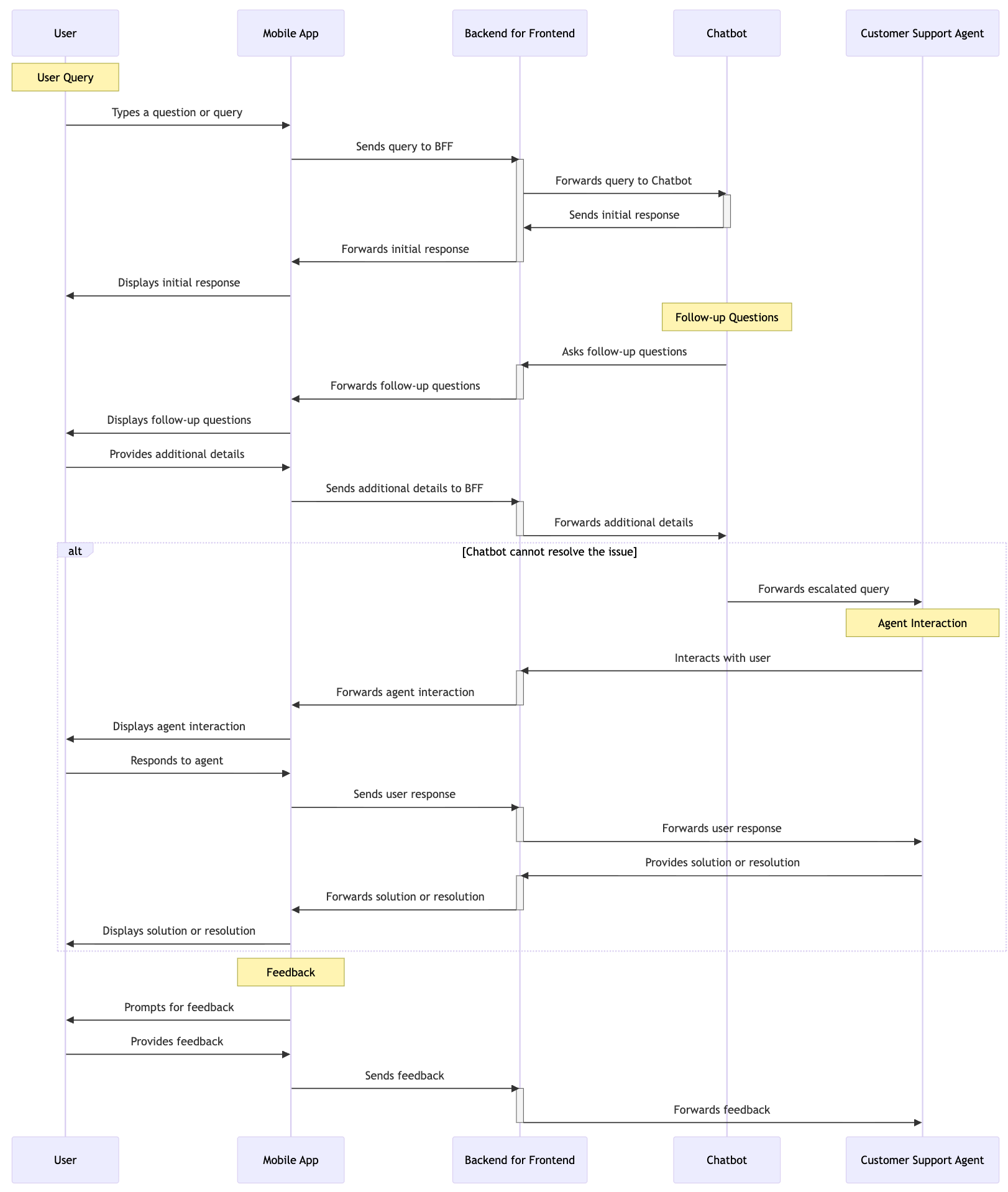}
    \caption{A UML sequence diagram from our benchmark, illustrating the ``Chatbot Interaction for Customer Support'' flow. Participants include the User, Mobile App, Backend For Frontend, Chatbot, and Customer Support Agent. In this scenario, the User submits a query through the Mobile App, which forwards it via the BFF to the Chatbot. The Chatbot provides initial responses and may request clarifications; if unable to resolve the issue, it escalates the conversation to a Customer Support Agent. The Agent then interacts with the User through the Mobile App to provide a resolution, after which the app collects feedback from the User.}
    \label{fig:uml-sequence-diagram-2}
\end{figure*}

Listing~\ref{lst:uml-sequence-nl} presents the natural language specification for the ``Uploading Documents with Secure Storage'' scenario (corresponding to the syntax provided in Listing~\ref{lst:uml-sequence-mermaid} and rendered in Figure~\ref{fig:uml-sequence-diagram}), while Listing~\ref{lst:uml-sequence-nl-2} provides the specification for the ``Chatbot Interaction for Customer Support'' flow (corresponding to the syntax provided in Listing~\ref{lst:uml-sequence-mermaid-2} and rendered in Figure~\ref{fig:uml-sequence-diagram-2}).


\begin{figure}[htbp]
\centering
\begin{lstlisting}[
    label=lst:uml-sequence-mermaid,
    basicstyle=\ttfamily\scriptsize,
    frame=single,
    caption={Mermaid syntax illustrating the ``Uploading Documents with Secure Storage'' flow, as rendered and detailed in Figure~\ref{fig:uml-sequence-diagram}.},
    captionpos=b,
    numbers=none,
    breaklines=true,
    breakindent=0pt,
    showstringspaces=false,
    linewidth=\columnwidth,
    xleftmargin=0.02\columnwidth,
    xrightmargin=0.02\columnwidth
]
sequenceDiagram
participant U as User
participant MA as Mobile App
participant BFF as Backend for Frontend
participant AAD as Azure AD
participant DB as Database
participant AS as Azure Blob Storage

Note over U: Upload Document
U->>MA: Upload document
MA->>BFF: Send document (with session token)
activate BFF
BFF->>AAD: Validate session token
deactivate BFF
activate AAD
AAD-->>BFF: Token validated
deactivate AAD
activate BFF
BFF->>AAD: Check user permissions
deactivate BFF
activate AAD
AAD-->>BFF: Permissions valid
deactivate AAD
activate BFF
alt Invalid token, insufficient permissions, or file too large
  BFF-->>MA: Error message
else Upload successful
  BFF->>DB: Save document metadata
  activate DB
  DB-->>BFF: Metadata saved
  deactivate DB
  BFF->>AS: Store document
  activate AS
  AS-->>BFF: Document stored
  deactivate AS
  BFF-->>MA: Confirmation
end
deactivate BFF
\end{lstlisting}
\end{figure}

\begin{figure}[htbp]
\centering
\begin{lstlisting}[
    label=lst:uml-sequence-mermaid-2,
    basicstyle=\ttfamily\scriptsize,
    frame=single,
    caption={Mermaid syntax illustrating the ``Chatbot Interaction for Customer Support'' flow, as rendered and detailed in Figure~\ref{fig:uml-sequence-diagram-2}.},
    captionpos=b,
    numbers=none,
    breaklines=true,
    breakindent=0pt,
    showstringspaces=false,
    linewidth=\columnwidth,
    xleftmargin=0.02\columnwidth,
    xrightmargin=0.02\columnwidth
]
sequenceDiagram
participant U as User
participant MA as Mobile App
participant BFF as Backend for Frontend
participant CB as Chatbot
participant CSA as Customer Support Agent

Note over U: User Query
U->>MA: Types a question or query
MA->>BFF: Sends query to BFF
activate BFF
BFF->>CB: Forwards query to Chatbot
activate CB
CB->>BFF: Sends initial response
deactivate CB
BFF->>MA: Forwards initial response
deactivate BFF
MA->>U: Displays initial response
Note over CB: Follow-up Questions
CB->>BFF: Asks follow-up questions
activate BFF
BFF->>MA: Forwards follow-up questions
deactivate BFF
MA->>U: Displays follow-up questions
U->>MA: Provides additional details
MA->>BFF: Sends additional details to BFF
activate BFF
BFF->>CB: Forwards additional details
deactivate BFF
alt Chatbot cannot resolve the issue
    CB->>CSA: Forwards escalated query
    Note over CSA: Agent Interaction
    CSA->>BFF: Interacts with user
    activate BFF
    BFF->>MA: Forwards agent interaction
    deactivate BFF
    MA->>U: Displays agent interaction
    U->>MA: Responds to agent
    MA->>BFF: Sends user response
    activate BFF
    BFF->>CSA: Forwards user response
    deactivate BFF
    CSA->>BFF: Provides solution or resolution
    activate BFF
    BFF->>MA: Forwards solution or resolution
    deactivate BFF
    MA->>U: Displays solution or resolution
end
Note over MA: Feedback
MA->>U: Prompts for feedback
U->>MA: Provides feedback
MA->>BFF: Sends feedback
activate BFF
BFF->>CSA: Forwards feedback
deactivate BFF
\end{lstlisting}
\end{figure}


\begin{figure}[htbp]
\centering
\begin{lstlisting}[
    label=lst:uml-sequence-nl,
    basicstyle=\ttfamily\scriptsize,
    frame=single,
    caption={Natural language specification for the ``Uploading Documents with Secure Storage'' flow, corresponding to the syntax in Listing~\ref{lst:uml-sequence-mermaid} and the rendered diagram in Figure~\ref{fig:uml-sequence-diagram}.},
    captionpos=b,
    numbers=none,
    breaklines=true,
    breakindent=0pt,
    showstringspaces=false,
    linewidth=\columnwidth,
    xleftmargin=0.02\columnwidth,
    xrightmargin=0.02\columnwidth
]
Purpose: Uploading Documents with Secure Storage

Main Components: User, Mobile App, BFF, Azure AD, Database

Interactions:
1. User Action: User uploads a document (e.g., an ID or contract) through the mobile app.
2. Mobile App: Sends the document along with the session token to the BFF.
3. BFF Validation:
Validates the session token with Azure AD.
Checks user permissions to ensure they are authorized to upload documents.
4. Storage Process:
The BFF saves metadata about the document (e.g., file name, size, upload timestamp) in the database.
The actual document is securely stored in cloud storage (e.g., Azure Blob Storage).
5. Response:
On successful upload, the BFF returns a confirmation to the app.
If the user is unauthorized or the file exceeds size limits, an appropriate error is returned.
\end{lstlisting}
\end{figure}

\begin{figure}[htbp]
\centering
\begin{lstlisting}[
    label=lst:uml-sequence-nl-2,
    basicstyle=\ttfamily\scriptsize,
    frame=single,
    caption={Natural language specification for the ``Chatbot Interaction for Customer Support'' flow, corresponding to the syntax in Listing~\ref{lst:uml-sequence-mermaid-2} and the rendered diagram in Figure~\ref{fig:uml-sequence-diagram-2}.},
    captionpos=b,
    numbers=none,
    breaklines=true,
    breakindent=0pt,
    showstringspaces=false,
    linewidth=\columnwidth,
    xleftmargin=0.02\columnwidth,
    xrightmargin=0.02\columnwidth
]
Purpose: Chatbot interaction for customer support

Main Components: User, Mobile App, Chatbot, Customer Support Agent, BFF

Interactions:
1. User Query: The user types a question or query into the mobile app.
2. Chatbot Engagement: The mobile app sends the query to the Chatbot via the BFF.
3. Initial Response: The Chatbot processes the query and sends an initial response to the mobile app.
4. Follow-up Questions: The Chatbot may ask follow-up questions to better understand the user's issue.
5. Escalation: If the Chatbot cannot resolve the issue, it escalates the query to a Customer Support Agent.
6. Agent Interaction: The Customer Support Agent receives the escalated query and interacts with the user through the mobile app.
7. Resolution: The Customer Support Agent provides a solution or resolution to the user's issue.
8. Feedback: The mobile app prompts the user to provide feedback on the support experience.
\end{lstlisting}
\end{figure}


\clearpage

\section{LLM-as-a-Judge Prompts}
\label{sec:judge-prompts}

We list the prompts used by the judge models to evaluate across the six criteria:
\begin{itemize}
    \item \textit{Syntax} (Listing~\ref{lst:judge-syntax})
    \item \textit{Mermaid Only} (Listing~\ref{lst:judge-mermaid-only})
    \item \textit{Logic} (Listing~\ref{lst:judge-logic})
    \item \textit{Completeness} (Listing~\ref{lst:judge-completeness})
    \item \textit{Activation Handling} (Listing~\ref{lst:judge-activation-handling})
    \item \textit{Error \& Status Tacking} (Listing~\ref{lst:judge-error-status-tracking})
\end{itemize}


\begin{figure}[htbp]
\centering
\begin{lstlisting}[
    label=lst:judge-syntax,
    basicstyle=\ttfamily\scriptsize,
    frame=single,
    caption={LLM-as-a-Judge prompt for evaluating MermaidJS syntax and structural correctness.},
    captionpos=b,
    numbers=none,
    breaklines=true,
    breakindent=0pt,
    showstringspaces=false,
    linewidth=\columnwidth,
    xleftmargin=0.02\columnwidth,
    xrightmargin=0.02\columnwidth
]
Evaluate the MermaidJS output under
<AGENT_RESPONSE> for syntax and structural correctness based on MermaidJS rules.
<EXPECTED_AGENT_RESPONSE> is provided below, treat it as one accepted solution reference; do not require a verbatim match.
#####
<AGENT_RESPONSE>
{agent_response}
#####
<EXPECTED_AGENT_RESPONSE>
{expected_agent_response}
#####
Is it proper MermaidJS syntax?
Are all participants declared using `participant ActorName` syntax?
Are activation/deactivation statements used and properly balanced?
Are `alt`, `else`, and `end` blocks closed correctly and nested if needed?
#####
---
Provide a numerical score (0.000 to 1.000) and a concise explanation.
Format the output as: <score>; <explanation>
Scoring scale:
- 0.000 to 0.200: Very poor;
- 0.201 to 0.400: Poor;
- 0.401 to 0.600: Fair;
- 0.601 to 0.800: Good;
- 0.801 to 0.999: Very good;
- 1.000: Perfect;
\end{lstlisting}
\end{figure}


\begin{figure}[htbp]
\centering
\begin{lstlisting}[
    label=lst:judge-mermaid-only,
    basicstyle=\ttfamily\scriptsize,
    frame=single,
    caption={LLM-as-a-Judge prompt for evaluating on strictly containing MermaidJS code only.},
    captionpos=b,
    numbers=none,
    breaklines=true,
    breakindent=0pt,
    showstringspaces=false,
    linewidth=\columnwidth,
    xleftmargin=0.02\columnwidth,
    xrightmargin=0.02\columnwidth
]
Evaluate the agent's output under <AGENT_RESPONSE> to ensure it strictly contains MermaidJS code only.
<EXPECTED_AGENT_RESPONSE> is provided below, treat it as one accepted solution reference; do not require a verbatim match.
#####
<AGENT_RESPONSE>
{agent_response}
#####
<EXPECTED_AGENT_RESPONSE>
{expected_agent_response}
#####
Is the output wrapped in a valid Markdown block (e.g., ```sequenceDiagram)?
Does it avoid extra explanation, narration, or formatting beyond valid Mermaid syntax?
Is the output clean, parsable, and renderable in Mermaid?
Responses with non-code text or mixed formatting should receive a lower score.
#####
---
Provide a numerical score (0.000 to 1.000) and a concise explanation.
Format the output as: <score>; <explanation>
Scoring scale:
- 0.000 to 0.200: Very poor;
- 0.201 to 0.400: Poor;
- 0.401 to 0.600: Fair;
- 0.601 to 0.800: Good;
- 0.801 to 0.999: Very good;
- 1.000: Perfect;
\end{lstlisting}
\end{figure}


\begin{figure}[htbp]
\centering
\begin{lstlisting}[
    label=lst:judge-logic,
    basicstyle=\ttfamily\scriptsize,
    frame=single,
    caption={LLM-as-a-Judge prompt for evaluating MermaidJS logic and flow completeness.},
    captionpos=b,
    numbers=none,
    breaklines=true,
    breakindent=0pt,
    showstringspaces=false,
    linewidth=\columnwidth,
    xleftmargin=0.02\columnwidth,
    xrightmargin=0.02\columnwidth
]
Evaluate the MermaidJS output under <AGENT_RESPONSE> based on the task description in <AGENT_PROMPT> (and, the reference in <EXPECTED_AGENT_RESPONSE>) for logic and flow completeness. The expected output is one accepted solution; alternate but logically equivalent flows should not be penalized.
#####
<AGENT_PROMPT>
{agent_prompt}
#####
<AGENT_RESPONSE>
{agent_response}
#####
<EXPECTED_AGENT_RESPONSE>
{expected_agent_response}
#####
Is it proper MermaidJS syntax?
Does every request have a corresponding response?
Are alternate flows (e.g., success/failure, if/else) represented completely and clearly?
Are nested decision branches handled as required?
Does the diagram account for every described interaction and path (or match the accepted reference flow)?
#####
---
Provide a numerical score (0.000 to 1.000) and a concise explanation.
Format the output as: <score>; <explanation>
Scoring scale:
- 0.000 to 0.200: Very poor;
- 0.201 to 0.400: Poor;
- 0.401 to 0.600: Fair;
- 0.601 to 0.800: Good;
- 0.801 to 0.999: Very good;
- 1.000: Perfect;
\end{lstlisting}
\end{figure}


\begin{figure}[htbp]
\centering
\begin{lstlisting}[
    label=lst:judge-completeness,
    basicstyle=\ttfamily\scriptsize,
    frame=single,
    caption={LLM-as-a-Judge prompt for evaluating on MermaidJS completeness.},
    captionpos=b,
    numbers=none,
    breaklines=true,
    breakindent=0pt,
    showstringspaces=false,
    linewidth=\columnwidth,
    xleftmargin=0.02\columnwidth,
    xrightmargin=0.02\columnwidth
]
Evaluate the MermaidJS output under <AGENT_RESPONSE> based on <AGENT_PROMPT> for completeness. <EXPECTED_AGENT_RESPONSE> is provided below, treat it as one accepted solution reference; do not require a verbatim match.
#####
<AGENT_PROMPT>
{agent_prompt}
#####
<AGENT_RESPONSE>
{agent_response}
#####
<EXPECTED_AGENT_RESPONSE>
{expected_agent_response}
#####
Is it proper MermaidJS syntax?
Does the diagram cover all participants, request/response pairs, and decision points described in the prompt (or equivalently as in the accepted reference)?
Are alternate flows and error paths handled as required?
Does the output reflect full coverage of described behavior, including minor flows?
#####
---
Provide a numerical score (0.000 to 1.000) and a concise explanation.
Format the output as: <score>; <explanation>
Scoring scale:
- 0.000 to 0.200: Very poor;
- 0.201 to 0.400: Poor;
- 0.401 to 0.600: Fair;
- 0.601 to 0.800: Good;
- 0.801 to 0.999: Very good;
- 1.000: Perfect;
\end{lstlisting}
\end{figure}


\begin{figure}[htbp]
\centering
\begin{lstlisting}[
    label=lst:judge-activation-handling,
    basicstyle=\ttfamily\scriptsize,
    frame=single,
    caption={LLM-as-a-Judge prompt for evaluating on MermaidJS activation handling.},
    captionpos=b,
    numbers=none,
    breaklines=true,
    breakindent=0pt,
    showstringspaces=false,
    linewidth=\columnwidth,
    xleftmargin=0.02\columnwidth,
    xrightmargin=0.02\columnwidth
]
Evaluate the use of activation and deactivation in the MermaidJS under <AGENT_RESPONSE> based on the <AGENT_PROMPT>. <EXPECTED_AGENT_RESPONSE> is provided below, treat it as one accepted solution reference; do not require a verbatim match.
#####
<AGENT_PROMPT>
{agent_prompt}
#####
<AGENT_RESPONSE>
{agent_response}
#####
<EXPECTED_AGENT_RESPONSE>
{expected_agent_response}
#####
Is it proper MermaidJS syntax?
Does the diagram properly use `activate` and `deactivate` to show control of execution?
Are all activated participants deactivated appropriately?
Are there any unnecessary `deactivate` statements where no activation occurred?
Does this improve the diagram's readability and traceability of actions?
#####
---
Provide a numerical score (0.000 to 1.000) and a concise explanation.
Format the output as: <score>; <explanation>
Scoring scale:
- 0.000 to 0.200: Very poor;
- 0.201 to 0.400: Poor;
- 0.401 to 0.600: Fair;
- 0.601 to 0.800: Good;
- 0.801 to 0.999: Very good;
- 1.000: Perfect;
\end{lstlisting}
\end{figure}


\begin{figure}[htbp]
\centering
\begin{lstlisting}[
    label=lst:judge-error-status-tracking,
    basicstyle=\ttfamily\scriptsize,
    frame=single,
    caption={LLM-as-a-Judge prompt for evaluating on MermaidJS error and status tracking.},
    captionpos=b,
    numbers=none,
    breaklines=true,
    breakindent=0pt,
    showstringspaces=false,
    linewidth=\columnwidth,
    xleftmargin=0.02\columnwidth,
    xrightmargin=0.02\columnwidth
]
Evaluate how clearly the MermaidJS under <AGENT_RESPONSE> handles error cases and status updates, based on the <AGENT_PROMPT>. <EXPECTED_AGENT_RESPONSE> is provided below, treat it as one accepted solution reference; do not require a verbatim match.
#####
<AGENT_PROMPT>
{agent_prompt}
#####
<AGENT_RESPONSE>
{agent_response}
#####
<EXPECTED_AGENT_RESPONSE>
{expected_agent_response}
#####
Is it proper MermaidJS syntax?
Does the diagram include explicit status updates?
Does the diagram clearly separate success and failure flows?
Does the diagram represent error-handling cases effectively?
Does the diagram track the state of key entities throughout the sequence?
#####
---
Provide a numerical score (0.000 to 1.000) and a concise explanation.
Format the output as: <score>; <explanation>
Scoring scale:
- 0.000 to 0.200: Very poor;
- 0.201 to 0.400: Poor;
- 0.401 to 0.600: Fair;
- 0.601 to 0.800: Good;
- 0.801 to 0.999: Very good;
- 1.000: Perfect;
\end{lstlisting}
\end{figure}

\clearpage
\section*{NeurIPS Paper Checklist}

\begin{enumerate}

\item {\bf Claims}
    \item[] Question: Do the main claims made in the abstract and introduction accurately reflect the paper's contributions and scope?
    \item[] Answer: \answerYes{} 
    \item[] Justification: The abstract and introduction accurately describe the benchmark's construction, evaluation method, and experimental scope. 

\item {\bf Limitations}
    \item[] Question: Does the paper discuss the limitations of the work performed by the authors?
    \item[] Answer: \answerYes{} 
    \item[] Justification: The paper includes a dedicated limitations section dsicussing the dataset size, syntactic coverage gaps, evaluation scope, and inherent biases of LLM-as-a-Judge evaluation.

\item {\bf Theory assumptions and proofs}
    \item[] Question: For each theoretical result, does the paper provide the full set of assumptions and a complete (and correct) proof?
    \item[] Answer: \answerNA{} 
    \item[] Justification: The paper contains no theoretical results or proofs as all contributions are empirical.

    \item {\bf Experimental result reproducibility}
    \item[] Question: Does the paper fully disclose all the information needed to reproduce the main experimental results of the paper to the extent that it affects the main claims and/or conclusions of the paper (regardless of whether the code and data are provided or not)?
    \item[] Answer: \answerYes{} 
    \item[] Justification: The dataset is publicly available on HuggingFace and the evaluation code is full open-source on GitHub. Inference parameters, all model identifiers, and full judge prompts are specified in the paper and appendix.

\item {\bf Open access to data and code}
    \item[] Question: Does the paper provide open access to the data and code, with sufficient instructions to faithfully reproduce the main experimental results, as described in supplemental material?
    \item[] Answer: \answerYes{} 
    \item[] Justification: The benchmark dataset is publicly available on HuggingFace and the evaluation code is full open-source on GitHub. Both are linked in the introduction.

\item {\bf Experimental setting/details}
    \item[] Question: Does the paper specify all the training and test details (e.g., data splits, hyperparameters, how they were chosen, type of optimizer) necessary to understand the results?
    \item[] Answer: \answerYes{} 
    \item[] Justification: All inference details are specified in the paper and full judge prompts are provided in the appendix. No training or optimization is involved.

\item {\bf Experiment statistical significance}
    \item[] Question: Does the paper report error bars suitably and correctly defined or other appropriate information about the statistical significance of the experiments?
    \item[] Answer: \answerNo{} 
    \item[] Justification: Error bars are not reported as all model outputs are generated deterministically using greedy decoding which eliminates run-to-run variance. Cross-judge consistency across two independent judge models serves as the primary robustness check.

\item {\bf Experiments compute resources}
    \item[] Question: For each experiment, does the paper provide sufficient information on the computer resources (type of compute workers, memory, time of execution) needed to reproduce the experiments?
    \item[] Answer: \answerNo{} 
    \item[] Justification: Compute resources are not reported as this work involves inference-only evaluations with no training. The computational cost is modest and entirely dependent on the user's chosen hardware.
    
\item {\bf Code of ethics}
    \item[] Question: Does the research conducted in the paper conform, in every respect, with the NeurIPS Code of Ethics \url{https://neurips.cc/public/EthicsGuidelines}?
    \item[] Answer: \answerYes{} 
    \item[] Justification: The research involves LLM evaluation for structured diagram generation on a synthetically constructed benchmark with no sensitive data, human subjects, or dual-use concerns.

\item {\bf Broader impacts}
    \item[] Question: Does the paper discuss both potential positive societal impacts and negative societal impacts of the work performed?
    \item[] Answer: \answerYes{} 
    \item[] Justification: Broader impacts are discussed in the Limitations and Broader Impacts section. The work contributes an open benchmark advancing rigorous evaluation practices with no meaningful negative societal impacts.
    
\item {\bf Safeguards}
    \item[] Question: Does the paper describe safeguards that have been put in place for responsible release of data or models that have a high risk for misuse (e.g., pre-trained language models, image generators, or scraped datasets)?
    \item[] Answer: \answerNA{} 
    \item[] Justification: The released dataset consists of synthetically constructed Mermaid sequence diagrams and poses no risk for misuse. No models were trained or released for this work.

\item {\bf Licenses for existing assets}
    \item[] Question: Are the creators or original owners of assets (e.g., code, data, models), used in the paper, properly credited and are the license and terms of use explicitly mentioned and properly respected?
    \item[] Answer: \answerYes{} 
    \item[] Justification: All models, tools, and datasets used in this paper are properly cited.

\item {\bf New assets}
    \item[] Question: Are new assets introduced in the paper well documented and is the documentation provided alongside the assets?
    \item[] Answer: \answerYes{} 
    \item[] Justification: The benchmark dataset and evaluation code are publicly released and open-source. The dataset construction, evaluation criteria, and judge prompts are fully documented in the paper and appendix.

\item {\bf Crowdsourcing and research with human subjects}
    \item[] Question: For crowdsourcing experiments and research with human subjects, does the paper include the full text of instructions given to participants and screenshots, if applicable, as well as details about compensation (if any)? 
    \item[] Answer: \answerNA{} 
    \item[] Justification: The paper does not involve crowdsourcing or formal research with human subjects. Dataset verification was conducted by internal subject matter experts.

\item {\bf Institutional review board (IRB) approvals or equivalent for research with human subjects}
    \item[] Question: Does the paper describe potential risks incurred by study participants, whether such risks were disclosed to the subjects, and whether Institutional Review Board (IRB) approvals (or an equivalent approval/review based on the requirements of your country or institution) were obtained?
    \item[] Answer: \answerNA{} 
    \item[] Justification: The paper does not involve human subjects research and requires no IRB approval.

\item {\bf Declaration of LLM usage}
    \item[] Question: Does the paper describe the usage of LLMs if it is an important, original, or non-standard component of the core methods in this research? Note that if the LLM is used only for writing, editing, or formatting purposes and does \emph{not} impact the core methodology, scientific rigor, or originality of the research, declaration is not required.
    \item[] Answer: \answerYes{} 
    \item[] Justification: LLMs are central to the methodology as they are used for synthetic dataset expansion for creating the benchmark dataset and for serving as LLM-as-a-Judge evaluators. All usage is full described in the paper.

\end{enumerate}

\end{document}